\documentclass[10pt,twocolumn,floatfix,aps,pra,superscriptaddress,]{revtex4-1} 
\usepackage{amsmath,amssymb,amsthm,mathrsfs,amsfonts,dsfont,amstext} 
\usepackage{textcomp,pbox}
\usepackage[export]{adjustbox}
\usepackage{bm}
\usepackage{dcolumn,booktabs,url}
\usepackage[scaled]{helvet}
\usepackage{sansmath,gensymb}
\usepackage{tikz,graphicx,transparent,color}
\usepackage{multirow}
\usepackage[separate-uncertainty = true]{siunitx}
\usepackage{comment}
\usepackage{physics}
\usepackage{graphicx}

\newcommand{\red}[1]{\textcolor{black}{#1}}

\usepackage[colorlinks=true]{hyperref}

\graphicspath{{./figures/}}

\hypersetup{
     colorlinks   = true,
     citecolor    = red,
     linkcolor    = blue,
     urlcolor     = red     
}

\newcommand{\yso}{Y$_2$SiO$_5$}

\newcommand{\ybionisotope}[0]{$^{171}$Yb$^{3+}$}
\newcommand{\ybyso}[0]{Yb$^{3+}$:Y$_2$SiO$_5$}
\newcommand{\yb}[0]{Yb$^{3+}$}
\newcommand{\ybiso}[0]{$^{171}$Yb$^{3+}$:Y$_2$SiO$_5$}
\newcommand{\y}[0]{Y$^{3+}$}
\newcommand{\transition}{$^2$F$_{7/2}(0) \longleftrightarrow ^2$F$_{5/2}$(0)  }
\newcommand{\figref}[1]{\figurename{~\ref{#1}}}
\newcommand{\tabref}[1]{\tablename{~\ref{#1}}}
\newcommand{\gstate}{$^2$F$_{7/2}$}
\newcommand{\estate}{$^2$F$_{5/2}$}

\begin{document}

\newcommand{\TitleName}{Spectroscopic study of hyperfine properties in \ybiso{}}

\title{\TitleName}

\newcommand{\AffGeneve}{Groupe de Physique Appliqu\'{e}e, Universit\'e de Gen\`{e}ve, CH-1211 Gen\`{e}ve, Switzerland}
\newcommand{\AffParis}{Chimie ParisTech, PSL University, CNRS, Institut de Recherche de Chimie Paris, 75005 Paris, France }
\newcommand{\AffPariss}{ Sorbonne Universit\'e, Facult\'e des Sciences et Ing\'enierie, UFR 933, Paris, France}

\author{Alexey~Tiranov}
\affiliation{\AffGeneve{}}
\author{Antonio~Ortu}
\affiliation{\AffGeneve{}}
\author{Sacha~Welinski}
\affiliation{\AffParis{}}
\author{Alban~Ferrier}
\affiliation{\AffParis{}}
\affiliation{\AffPariss{}}
\author{Philippe~Goldner}
\affiliation{\AffParis{}}
\author{Nicolas~Gisin}
\affiliation{\AffGeneve{}}
\author{Mikael~Afzelius}\email[Email to: ]{mikael.afzelius@unige.ch}
\affiliation{\AffGeneve{}}

\date{\today}

\begin{abstract}
	Rare-earth ion doped crystals are promising systems for quantum communication and quantum information processing. In particular, paramagnetic rare-earth centres can be utilized to realize quantum coherent interfaces simultaneously for optical and microwave photons. In this article, we study hyperfine and magnetic properties of a \yso{} crystal doped with $^{171}$\yb{} ions. This isotope is particularly interesting since it is the only rare--earth ion having electronic spin $S=\frac{1}{2}$ and nuclear spin $I=\frac{1}{2}$, which results in the simplest possible hyperfine level structure. In this work we determine the hyperfine tensors for the ground and excited states on the optical \transition{} transition by combining spectral holeburning and optically detected magnetic resonance techniques. The resulting spin Hamiltonians correctly predict the magnetic-field dependence of all observed optical-hyperfine transitions, \red{from zero applied field up to fields where the Zeeman interaction is dominating the hyperfine interaction}. Using the optical absorption spectrum we can also determine the order of the hyperfine levels in both states. These results pave the way for realizing solid-state optical and microwave quantum memories based on a~\ybiso{} crystal.
\end{abstract}

\maketitle

\section{INTRODUCTION}

Rare-earth ion-doped crystals (REIC) are of particular interest in domains of quantum information processing and quantum communication~\cite{Tittel2010b,Thiel2011,Bussieres2013,Riedmatten2015,GoldnerChapter2015}. Thanks to their long optical and spin coherence times at low temperatures~\cite{Macfarlane2002,Bottger2009} and wide optical  inhomogeneous linewidths (in the GHz range) they have been actively studied to realize optical quantum memories~\cite{Tittel2010b,Bussieres2013} and quantum processors~\cite{Longdell2004,Ahlefeldt2013}. 
In this context the optically addressable hyperfine transitions of Eu$^{3+}$:\yso{} showed coherence lifetimes up to six hours~\cite{Arcangeli2014,Zhong2015}. Recent progress of optical memory experiments includes spin-wave storage~\cite{Jobez2015,Gundogan2015,Seri2017}, as well as memories with high efficiency~\cite{Sabooni2013,Jobez2014}, and long storage time~\cite{Heinze2013}. Quantum applications of REIC involve photonic entanglement storage~\cite{Saglamyurek2011,Clausen2011} of different types~\cite{Tiranov2015a,Tiranov2016b} and light-matter teleportation at a telecom wavelength~\cite{Bussieres2014}. The interface between single RE ions have also been demonstrated~\cite{Kolesov2012,Utikal2014}, which opens the way to quantum processing using these systems. 

RE ions with an odd number of electrons such as Nd$^{3+}$, Er$^{3+}$ and Yb$^{3+}$ are known to form paramagnetic $S=\frac{1}{2}$ centers (Kramers doublets) when doped into low-symmetry crystal sites~\cite{MacfarlaneShelby1987}, with magnetic moments of the order of the electronic Bohr magneton $\mu_B~\approx~14$~GHz/T. Due to this fact they can be interfaced with microwave photons through a superconducting resonator~\cite{Probst2013,Wolfowicz2015,Chen2016}. This approach gives additional tools for hybrid quantum technologies, since the nuclear hyperfine transitions for some istotopes can also provide long coherence time, as for instance more than one second in $^{167}$Er${^{3+}}$:\yso{}~\cite{Rancic2018}. The transfer of coherence between the electronic and nuclear spins with high-fidelity was achieved for $^{145}$Nd${^{3+}}$:\yso{} with nuclear coherence times up to 9~ms~\cite{Wolfowicz2015}. 

Ytterbium has a number of advantages comparing with other rare-earths. The [Xe]4f$^{13}$ electronic configuration of \yb{} results in a simple energy level structure consisting of only two electronic multiplets: $^2$F$_{7/2}$ and $^2$F$_{5/2}$ for ground and excited states, respectively. The optical line between the lowest energy levels of the ground and excited multiplets (around 980~nm) is easily accessible by standard diode lasers, while commercial single photon sources based on InGaAs quantum dots are also covering these optical energies~\cite{Loredo2016}. The favorable optical branching ratio connecting lowest crystal field levels, in comparison to other REIC systems,  gives additional advantages for the single ion detection using optical cavity enhancement~\cite{Ding2016,Miyazono2016}. Another advantage comes from the specific isotope $^{171}$Yb, which has the lowest non--zero nuclear spin ($I=\frac{1}{2}$). This fact greatly simplifies the energy level structure facilitating the spectral tailoring and the spectroscopic study using standard methods, as for instance it limits the number of different transition lines observed in absorption spectra, making them easier to identify.

The basic spectroscopic properties of a naturally doped \ybyso{} crystal were presented in Ref.~\cite{Welinski2016}. These included absorption coefficients as a function of polarization, characterization of radiative lifetimes and branching ratios, as well as magnetic properties of the ground and excited state. In this work we present a detailed spectroscopic study of the particular isotope \ybionisotope{} in the same \yso{} host, using spectral hole burning (SHB) and optically detected magnetic resonance (ODMR) techniques. We find that the \ybionisotope{} hyperfine tensor, deduced from standard electron paramagnetic resonance (EPR) measurements in Ref.~\cite{Welinski2016}, does not correctly predict the zero-field ODMR resonances, nor their magnetic field dependence in the intermediate non-linear field regime \red{where the hyperfine and Zeeman interactions have similar strength}. We attribute this to the problem of properly determining the hyperfine tensor from EPR data when the tensor is anisotropic and the doping site has low symmetry ($C_1$ in this case).

For the same reason, the spin Hamiltonian of $^{167}$Er$^{3+}$:\yso{} crystal was recently refined by combining the EPR measurements at low and high fields~\cite{Chen2018}. Using our SHB and ODMR data  together with previous EPR measurements \cite{Welinski2016} allows us to determine the spin Hamiltonian for both the ground and excited states of \ybiso{}. The predictions of the new spin Hamiltonians were compared with experimental data in a large range of magnetic fields, yielding an excellent agreement. Using the absorption profile measurements and selective ODMR we also determine the order of the energy levels on the optical \transition{} transition. 

\red{We note that throughout this article we call the \textit{low-field regime} where the hyperfine energy is larger than the hyperfine interaction, the \textit{intermediate regime} where they have similar energy, and the \textit{high-field regime} where the Zeeman energy is larger than the hyperfine energy. In any case we do not consider the regime where the Zeeman interaction is non-linear due to perturbations from higher-lying crystal field levels \cite{Mehta2000}.}

We additionally observe peculiar transformations of holes to antiholes in the SHB spectra, and vice versa, as the field intensity is varied across a nonlinear field regime. We interpret this as a change in spin cross-relaxation rates due to a transformation of the wavefunctions in the studied magnetic field region.

The article is organized as follows. In Section~\ref{sec:Ham} we briefly describe the spin Hamiltonian utilized for spin $\frac{1}{2}$ systems and our \ybiso{} crystal. In Section~\ref{sec:exp} we describe the setup and methods of SHB and ODMR. Section~\ref{sec:results} shows the main results: the measurement of ground state and excited state hyperfine splittings as a function of the external magnetic field and the refined spin Hamiltonian parameters.  In Section~\ref{sec:holes} we discuss the observed exchange of wavefunctions under an avoided crossing condition. We finally discuss the implications of our findings and give an outlook in Section~\ref{sec:conclusion}.

\section{BASIC PROPERTIES OF \ybiso{}}
\label{sec:Ham}

\subsection{Spin Hamiltonian}

Let us consider a system with both \red{electronic and nuclear spin 1/2 ($S = 1/2$ and $I = 1/2$)}. In this case, the electronic spin $\mathbf{S}$ is coupled with its nuclear spin $\mathbf{I}$ through the hyperfine interaction tensor~$\mathbf{A}$, and the effective spin Hamiltonian involving the interaction with an external magnetic field $\mathbf{B}$ can be written as~\cite{Abragam1970}
\begin{equation}
\label{eq:Heff}
\mathcal{H} = \mathbf{I} \cdot \mathbf{A} \cdot \mathbf{S} + \mu_\text{B} \mathbf{B} \cdot \mathbf{g} \cdot \mathbf{S} - \mu_\text{n}  \mathbf{B}\cdot \mathbf{g}_\text{n} \cdot \mathbf{I}.
\end{equation}
Here, $\mathbf{g}$ and $\mathbf{g}_\text{n}$ are the coupling tensors of the electronic and nuclear Zeeman interactions, respectively, while $\mu_B$ and $\mu_n$ are the electronic and nuclear magnetons. The quadrupolar interaction term $\mathbf{I}\cdot\mathbf{Q}\cdot\mathbf{I}$, present only for spin number $I\geq 1$ systems, does not appear for the $I=1/2$ of $^{171}$Yb. This strongly simplifies the spectroscopic analysis of crystalline systems doped with this isotope if compared to, for instance, $^{167}$Er$^{3+}$:\yso{}, which has $I=7/2$. \red{The nuclear Zeeman interaction is considered to be isotropic with $g_n= 0.987$. It is included for completeness, but it gives too small energy shifts to be detected for the range of magnetic fields used in this work.}

If there is no applied magnetic field ($\mathbf{B} = 0$), then the spin Hamiltonian can be diagonalized analytically, resulting in four states with energies
\begin{equation}
	\frac{1}{4} \left[ - A_3 \pm (A_1 + A_2) \right] \text{,  } \frac{1}{4} \left[ A_3 \pm (A_1 - A_2) \right],
	\label{eq:ZFlevels}
\end{equation}
\noindent where $A_1$, $A_2$ and $A_3$ are the eigenvalues of the $\mathbf{A}$ tensor. In the \yso{} crystal, $^{171}$Yb$^{3+}$ ions substitute Y$^{3+}$ ions in sites with $C_1$ point symmetry (see Sec. \ref{sec:crystal}). For this low symmetry the $\mathbf{A}$ tensor has three independent eigenvalues $A_1 \neq A_2 \neq A_3$, such that the zero-field hyperfine structure consists of four non-degenerate energy levels. If these are known, then the $\mathbf{A}$ tensor eigenvalues can be calculated analytically using Eq.~\eqref{eq:ZFlevels}. In Section \ref{sec:results} we use ODMR and SHB measurements to determine the zero-field energy level splittings, from which the tensor elements are calculated.

In $C_1$ point symmetry, the orientation of the hyperfine tensor with respect to the crystal axes is not given by the symmetry. Hence, to fully characterize the hyperfine tensor one also needs to determine three orientation angles. This can be done by applying a magnetic field ($\mathbf{B} \neq 0$), and studying the energy levels as a function of the field vector $\mathbf{B}$ (by varying its angle and/or strength). In this work we were also aided by the fact that the electronic Zeeman tensors $\mathbf{g}$ were fully known \cite{Welinski2016}, both for the ground and excited states. Therefore only the three orientation angles of the $\mathbf{A}$ were free parameters when fitting the field measurement data to the spin Hamiltonian, as we discuss in detail in Section~\ref{sec:results}.

In conventional EPR spectroscopy, all six independent elements of the $\mathbf{A}$ tensor (for $C_1$ site symmetry) are deduced from measurements with applied magnetic fields. When using EPR at the common 9.7~GHz X band, the energy splittings are mostly given by the electronic Zeeman interaction part in Eq.~\eqref{eq:Heff}, while the hyperfine interaction often has much less impact in this energy range. There is thus a question if X-band EPR data is sufficient to accurately fit all six elements in a $C_1$ symmetry. It should be emphasized that a hyperfine tensor $\mathbf{A}$ fitted using EPR data will generally be accurate in the range of fields where it was measured, as was the case in Ref.~\cite{Welinski2016}. However, as shown in this work, the accuracy of those hyperfine elements can be insufficient in order to predict the energy levels outside this measurement region. Particularly in the low-field regime, where the hyperfine interaction dominates, or in the intermediate field region where both the Zeeman and hyperfine interactions have similar strengths.

\begin{figure}
	\includegraphics[width=1\linewidth,trim=0.0cm -1.0cm 0cm 0cm,clip]{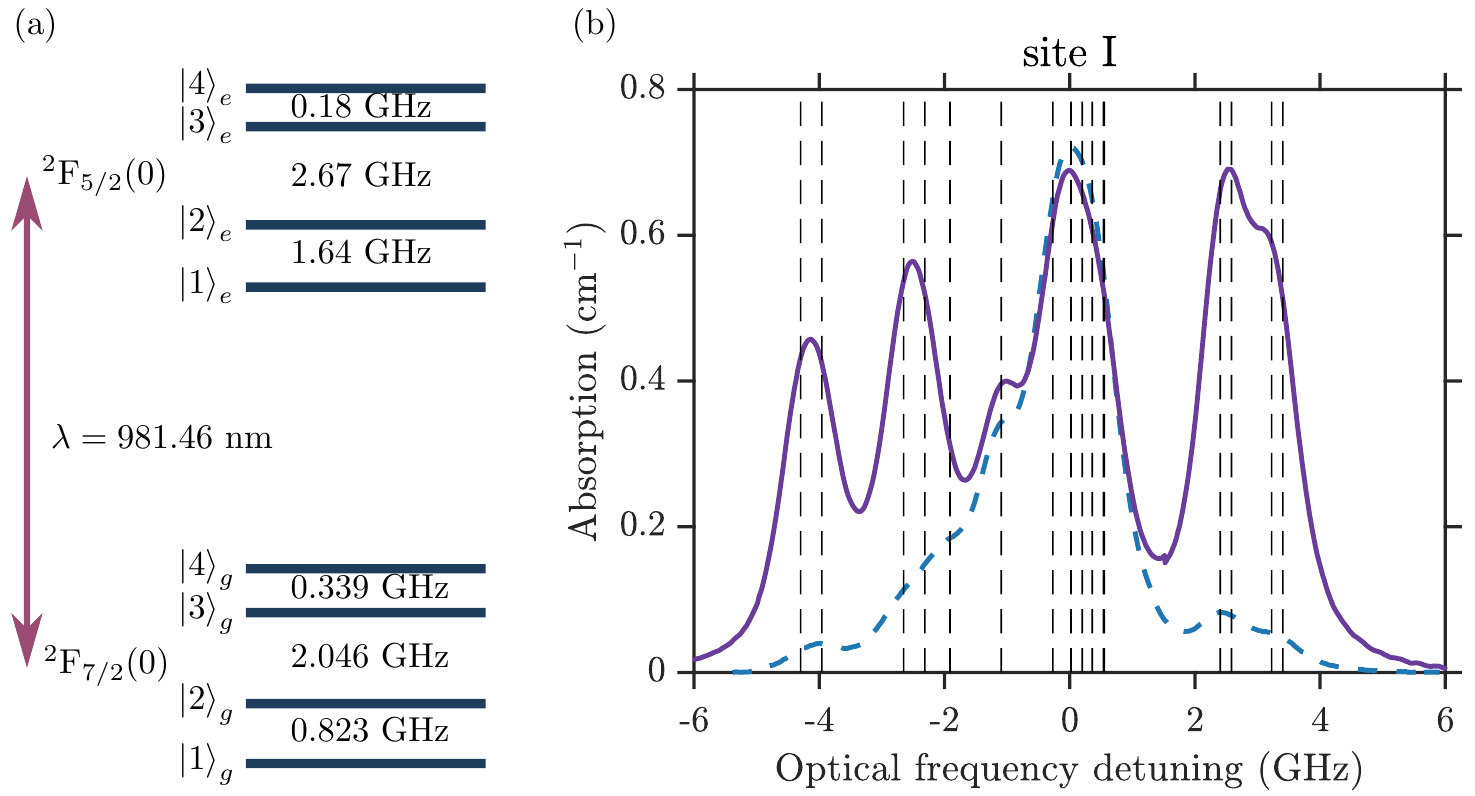}
	\includegraphics[width=1\linewidth,trim=0.0cm 0.0cm 0cm 0cm,clip]{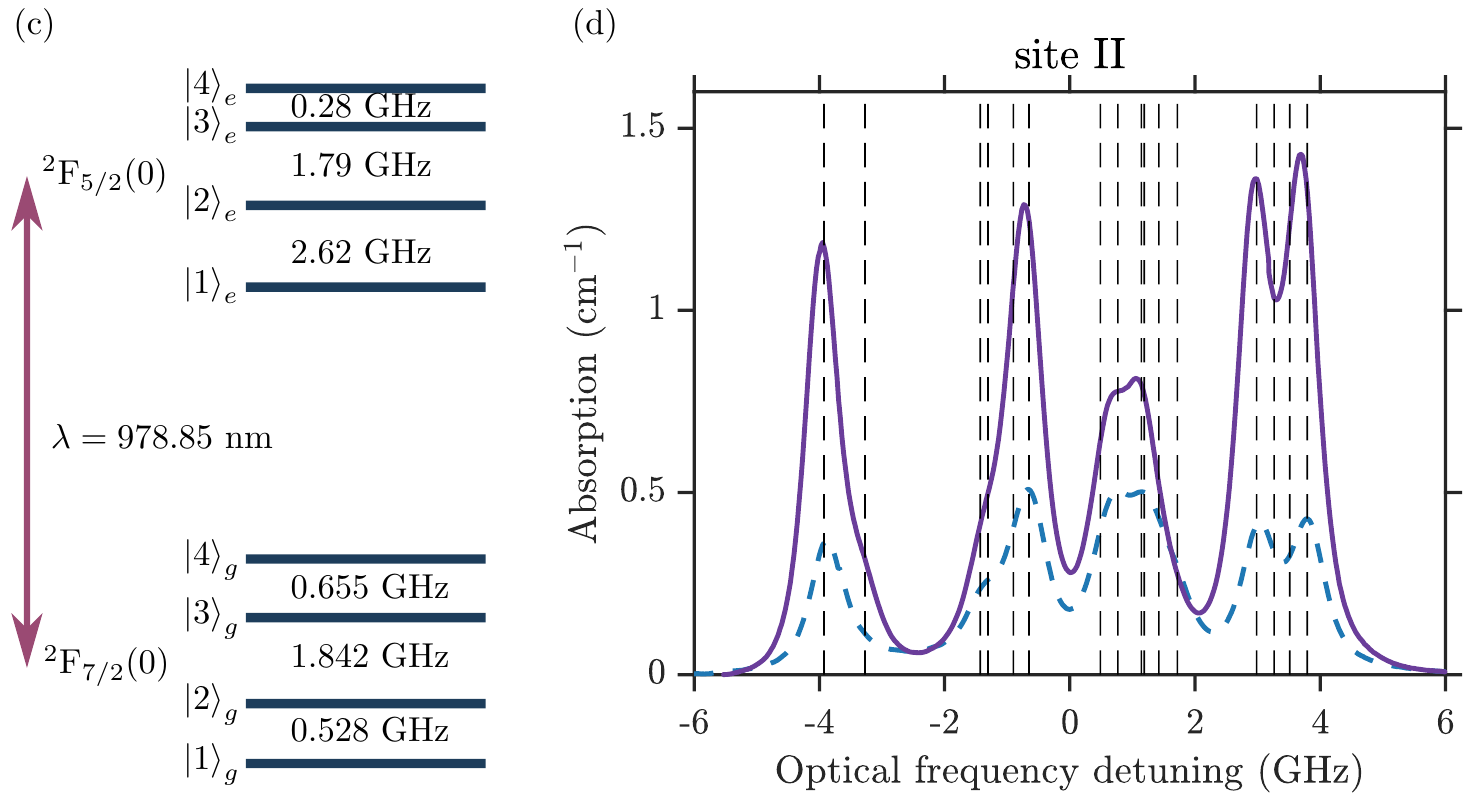}
	\caption[]{\label{fig:abs} (color online)
		\red{Energy level diagram of \ybiso{} showing the zero-field hyperfine level structures of the lowest crystal-field levels of the ground \gstate{}(0) and excited \estate{}(0) states, both for site~I (a) and site~II (c). The ground-state hyperfine level structures were determined from the zero-field ODMR measurements, while the excited-state structure was determined using SHB measurements (see Section~\ref{sec:results}). In (b) and (d) we show high-resolution optical absorption spectra of the \transition{} transition for site~I and site~II, respectively. These were recorded at low temperature (3~K) for light polarized along $\mathbf{D}_2$ (solid line) and $\mathbf{D}_1$ (dashed line) crystal axis. The position of all optical-hyperfine transitions have been calculated using the hyperfine $\mathbf{A}$ tensors measured in this work (dashed vertical lines). Zero optical detuning refers to the central transition wavelength given in Sec. \ref{sec:crystal}.}}
\end{figure}

\begin{figure*}
	\includegraphics[width=0.6\linewidth,trim=0.2cm 0.1cm 0.0cm 0.5cm,clip]{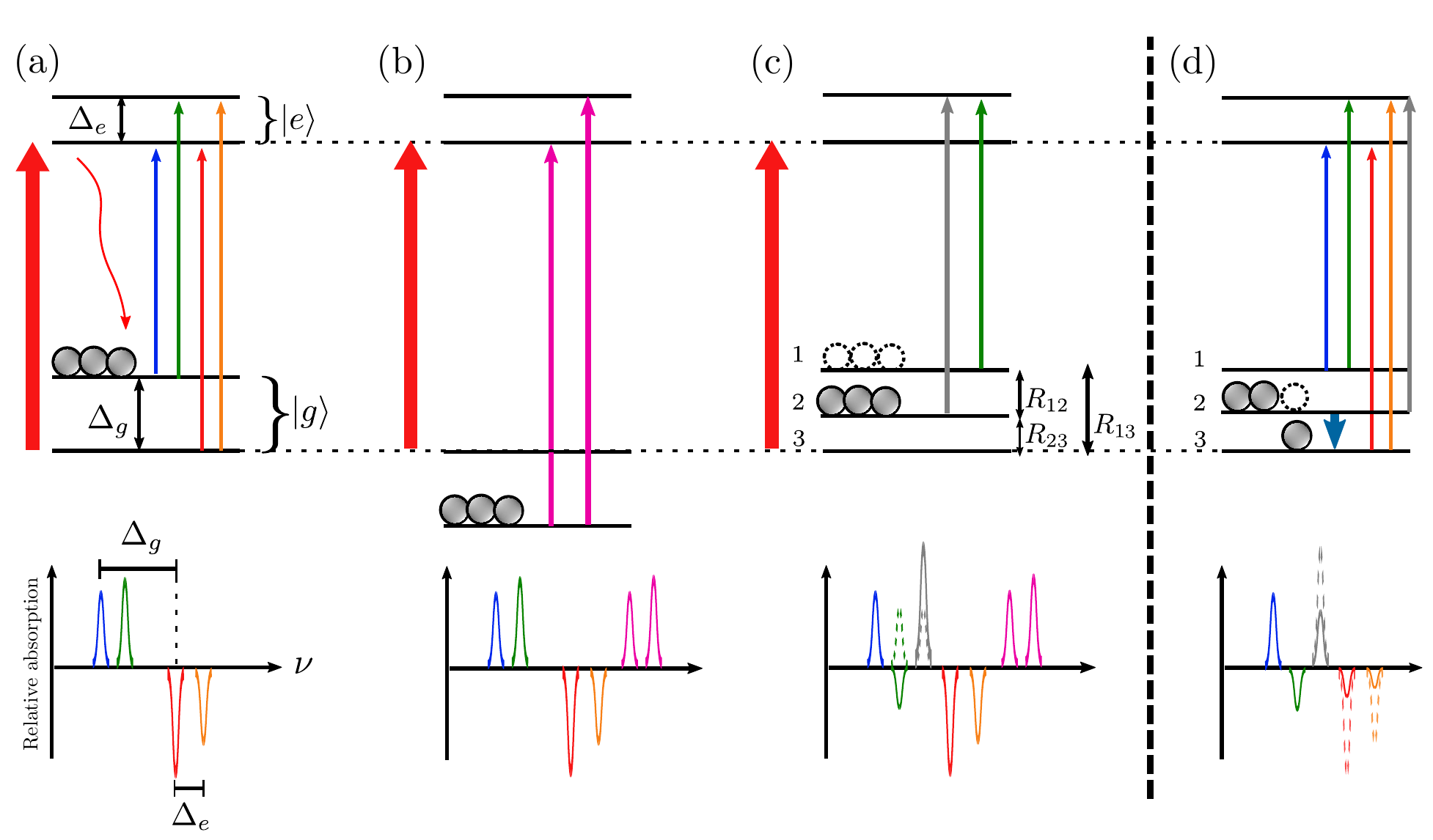}
	\hspace{1cm}
	\includegraphics[width=0.26\linewidth]{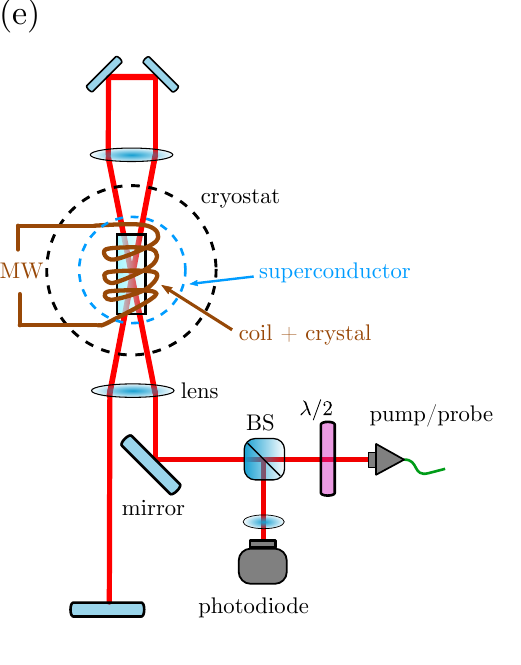}
	\caption{(color online) Examples of the SHB and ODMR techniques and experimental setup. (a) During a burning pulse (large red arrow), the pumped atoms are redistributed among the ground levels (wavy arrow). A weaker probe pulse is then sent on the system while the frequency is scanned, making holes and antiholes visible in the absorption spectrum (bottom). The splitting between two excited states $\Delta_e$ can be deduced from the difference in frequency between two holes with a common ground state. Similarly, ground splittings $\Delta_g$ can be found from the difference of frequencies related to antiholes with a common excited state. (b) Example of additional antiholes due to a different class of atoms resonant with the burn pulse in (a). (c) Example of transformation from antihole to hole when the ground state is made of more than two levels. If the decay rates connecting each upper ground level with the burned level ($R_{13}$ and $R_{23}$) are much lower than the rate connecting themselves $R_{12}$, the probe will produce antiholes when scanned over levels 1 and 2, which is the usual situation. However, population on level 1 can be depleted during burning if $R_{13}\gg R_{12},R_{23}$. Thus, a hole appears instead of the antihole seen in the previous case.
(d) Schematic explanation of ODMR for a microwave transition. If after the burning pulse all the pumped ions are in $2$, by applying microwave radiation resonant with the $2\leftrightarrow 3$ transition (thick dark blue arrow), the population will redistribute among the levels involved, leading to a decrease of the related holes/antihole structure. (e)~The experimental setup. See Section~\ref{sec:expsetup} for details.}
	\label{fig:setup_SHB}
\end{figure*}

\begin{figure*}
	\includegraphics[width=0.83\linewidth,trim=0.4cm 0.2cm 0cm 1.5cm,clip]{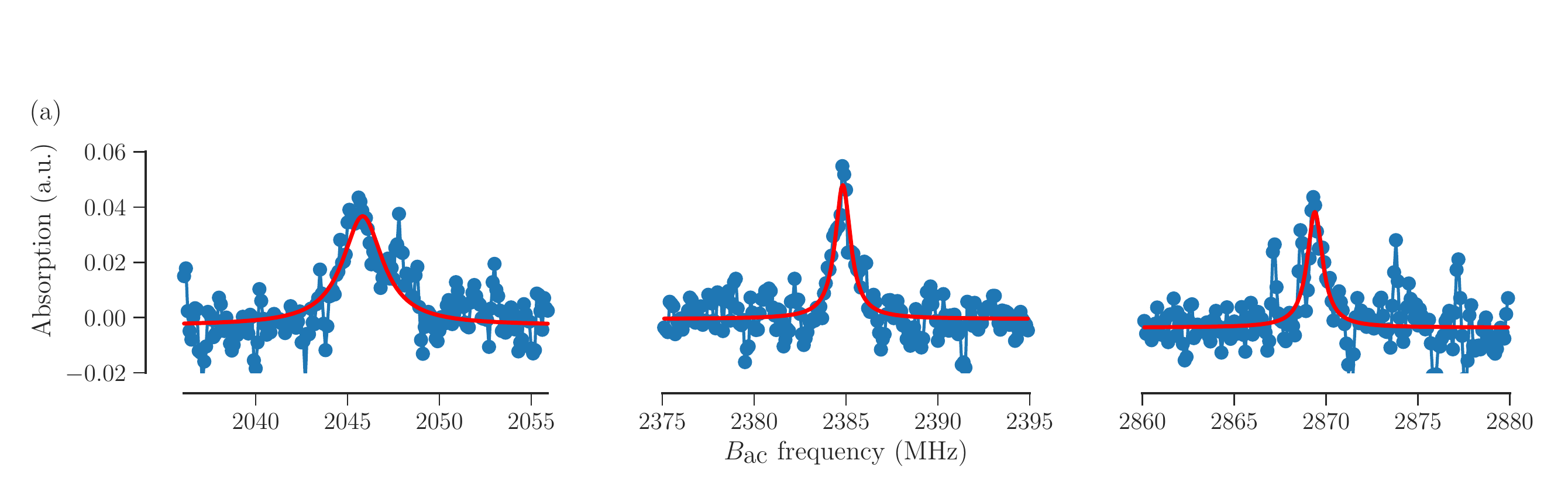}\includegraphics[width=0.17\linewidth,trim=0.0cm -0.9cm 0cm 0cm,clip]{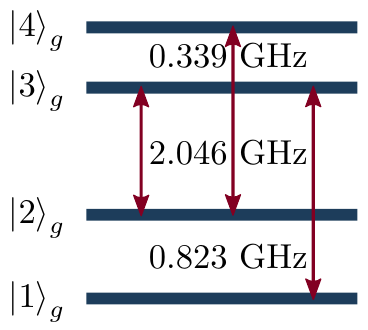}\\
	\includegraphics[width=1.0\linewidth,trim=3.1cm 0.5cm 4cm 0.8cm,clip]{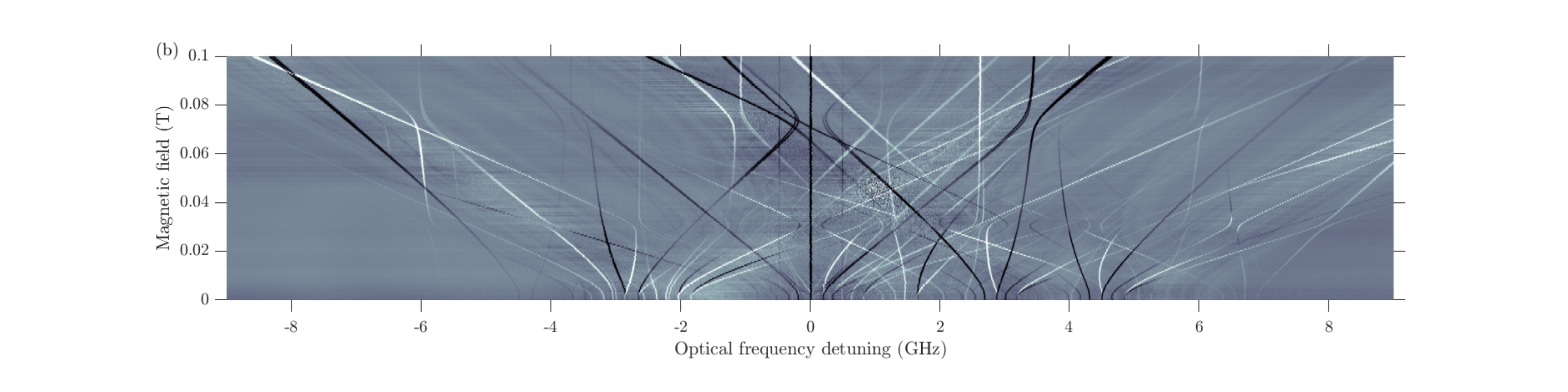} 
	\caption[]{\label{fig:map} (color online) Experimental results. (a) Optically detected magnetic resonance (ODMR) lines at zero magnetic field for site~I of \ybiso{}. The oscillating magnetic field $B_{\text{ac}}$ was applied in $\mathbf{b}$ direction. Three different transitions are shown and their corresponding levels are depicted on the right. The variation of the measured spin linewidths could be attributed to the power broadening effect.
(b) Recorded optical spectral holeburning (SHB) spectra of  site~I of \ybiso{} crystal measured for different magnetic field amplitudes applied in the direction close to $\mathbf{D}_1$-axis. Zero frequency detuning corresponds to the central frequency at which spectral holeburning is performed. Black regions correspond to lower absorption (holes), while white lines correspond to lower transmission (antiholes) regions. \red{The bending of some of the holes and anti-holes at around 30~mT of magnetic field is a result of avoided crossings of the associated ground-state levels. Many of these also show an unusual a hole-antihole transformation in this field region (see also \figref{fig:hole} and Sec. \ref{sec:holes}).}}
\end{figure*}

\subsection{Crystal properties and optical spectra}
\label{sec:crystal}
Our crystal is a \yso{} host doped with 10 ppm $^{171}$Yb$^{3+}$ with 95~\% isotope purity. It was grown via the Czochralski method and cut along the $\mathbf{D_1}$, $\mathbf{D_2}$ and $\mathbf{b}$ polarization extinction axes~\cite{Li1992}. The sides parallel to these axes have lengths $5.7$, $5.6$ and $\SI{9.5}{\milli\metre}$, 
and the faces corresponding to the $\mathbf{D_1}-\mathbf{D_2}$ plane were polished to reduce light scattering. \yso{} is a common crystal in the rare-earth community~\cite{Equall1994,Fraval2004,Bottger2006a,Zhong2015}, as it allows RE dopant ions to reach long coherence times due to the low nuclear spin density of the crystal. It has a monoclinic structure and it belongs to the $C_{2h}^6$ space group.

\yb{} ions can replace \y{} in the host crystal in two different sites of $C_1$ point symmetry, usually called site~I and site~II~\cite{Kurkin1980}. Furthermore, each crystallographic site consists of two magnetic sub-sites which are related to each other via the $C_2$ symmetry axis (this axis coincide with the $\mathbf{b}$  axis).

\red{The optical transition \transition{} under study here couples the lowest crystal field levels of the two electronic multiplets, with a vacuum wavelength of $\lambda=981.463$~nm for site I and $\lambda=978.854$~nm for site~II~\cite{Welinski2016}. The optical absorption spectra of this transition are shown in \figref{fig:abs}, which were recorded at zero magnetic field. Since both the ground and excited states have four non-degenerate energy levels, each spectrum consists of 16 optical-hyperfine transitions that are inhomogeneously broadened. The inhomogeneous profile was measured to be Lorentzian, with a full-width at half-maximum (FWHM) of 800~MHz for site~I and 560~MHz for site~II. Although the spectra are partly resolved, one cannot accurately measure the hyperfine splittings using the inhomogeneous absorption spectra. For this purpose we employ the SHB measurement technique.}

\section{EXPERIMENTAL METHODS}
\label{sec:exp}

\subsection{Spectral Hole Burning}
The SHB technique is commonly used to investigate transitions in systems with strong inhomogeneous broadening~\cite{MacFarlane1981}. In solid state crystals, inhomogeneous broadening is often due to uneven strains in different positions of the crystal~\cite{STONEHAM1969}, so that each ion can have a different detuning compared to the centre of the frequency distribution of a certain transition.

In this context, SHB consists in irradiating the sample with laser light at a specific frequency inside the broadened spectrum (the burn pulse), so to pump atoms away into other energy levels via an excited state. A scan of a weaker probe around the pump frequency will then reveal a series of holes and antiholes corresponding to transitions from states with increased or decreased population, respectively (\figref{fig:setup_SHB}).

\red{If the inhomogeneous broadening is much larger than the ground and excited state splittings, then a burn pulse at a fixed frequency can resonantly excite atoms in different \emph{classes}, where for each class the burn pulse excites a different optical-hyperfine transition \cite{MacFarlane1981,Hastings-Simon2008}. Each class will produce a pattern of holes and anti-holes, see \figref{fig:setup_SHB}, and the resulting spectrum can be hard to interpret. Since for this system the inhomogeneous broadening is smaller than some of the hyperfine splittings, the number of contributing classes will depend on the exact burn frequency. It also causes an asymmetric hole/anti-hole pattern, with respect to the burn frequency.} The hole/antihole structures can be studied as a function of the time between the pump and the probe pulses to reveal population lifetimes~\cite{Hastings-Simon2008}, and as a function of other external parameters \red{such as magnetic field and temperature~\cite{ZambriniCruzeiro2017a}}.

In this paper, we focus on the dependence of the position of spectral holes/antiholes on an external magnetic field, which induces shifts in frequency of the holes/antiholes as the energy level splittings are varied by Zeeman interaction\cite{Riesen2010}.

As will be explained in Section~\ref{sec:results}, we can also observe \textit{hole$\leftrightarrow$antihole} transformations as the magnetic field is varied across a critical value (\figref{fig:setup_SHB}c). This phenomenon marks the passage from a low field regime with highly nonlinear energy shifts due mostly to the hyperfine interaction, to a high field, linear regime dominated by Zeeman interactions.
\red{Hole/antihole transformations have been observed in different contexts including modification of the cross-relaxation \cite{Riesen2007} or temperature dependant phononic relaxation \cite{Hastings-Simon2008a}, or as a result of superhyperfine resonances between a thulium dopant and aluminium ions in a Tm$^{3+}$:YAG crystal~\cite{Ahlefeldt2015}.}

\subsection{Optically Detected Magnetic Resonance}
\label{sec:ODMR}
Magnetic resonance techniques can be combined with SHB to study transitions in domains different than the optical one~\cite{Laplane2016,Zhong2015,Fernandez-Gonzalvo2015,Macfarlane2014}. Hyperfine and Zeeman interactions often induce splittings in the microwave range in atomic ensembles, and these splittings can be varied by using an external magnetic field in a broad range between MHz and GHz frequencies, requiring a broadband investigation technique. We thus rely on ODMR to study transitions between spin states for various conditions of external magnetic field. The technique consists in burning a spectral hole and observing its variation as a microwave transition is addressed at the same time as an optical probe beam is on at a constant frequency (\figref{fig:setup_SHB}d)~\cite{MacfarlaneShelby1987}. When the microwave radiation is in resonance with the spin transition, the latter will be detected as a population change, which in turn reduces the amplitude of the spectral hole.

To find the spin resonance and obtain information such as its central frequency and linewidth, we measure the hole amplitude while scanning the microwave frequency. The measurements can be repeated at various external magnetic field values so to obtain information about its influence on the transition.

\subsection{Experimental setup}
\label{sec:expsetup}
Our setup is shown schematically in~\figref{fig:setup_SHB}e. The crystal is placed inside a cryostat in vacuum and at temperature of about $\SI{3}{\kelvin}$. A copper coil surrounds the crystal with its longitudinal axis coincident with the crystal $\mathbf{b}$ axis, and it is used in the ODMR technique to address the spin transitions by generating a microwave field (up to $\SI{4 }{\giga\hertz}$ in frequency, fed with power up to $\sim\SI{2 }{\watt}$).
A superconductor coil surrounds the crystal and copper coil, and allows us to apply static fields up to $\SI{2}{\tesla}$ in directions orthogonal to $\mathbf{b}$. In the experiments presented here, we use a pulsed laser beam (through an acousto-optic modulator, not shown), generated by a tunable $\sim\SI{980}{\nano\metre}$ external cavity diode laser and injected into a single mode polarization maintaining fibre. A half-waveplate right after the fibre output allows the adjustment of polarization according to the orientation of the crystal extinction axis of maximal absorption ($\mathbf{D}_2$ axis). The light is then focused on the sample in a spot with a diameter of $\sim\SI{100}{\micro\metre}$ and traverses it in four passes to further increase the amount of light absorbed. Finally, a beam splitter redirects the output light on a silicon photodiode.

\section{Experimental results}
\label{sec:results}

In principle the SHB measurement can provide all required information about energy splittings for the ground and excited states. However, the interpretation of the SHB data is difficult as most anti-holes depend on both the ground and excited state splittings. Moreover, as explained in~\figref{fig:setup_SHB}c, one can observe so-called pseudoholes, where one would expect to see an anti-hole, further complicating the interpretation.

To facilitate the analysis we first detect the hyperfine splittings of the ground state at zero magnetic field, using the ODMR technique explained previously (see Section~\ref{sec:ODMR}). For this purpose the laser producing the burn and probe pulses  was tuned to the middle of the absorption structure (\figref{fig:abs}) in order to address a maximum number of classes and to produce a population difference for a high number of spin transitions. 

\begin{figure*}
\includegraphics[width=0.93\linewidth,trim=2.0cm 0.cm 2.0cm 0.05cm,clip]{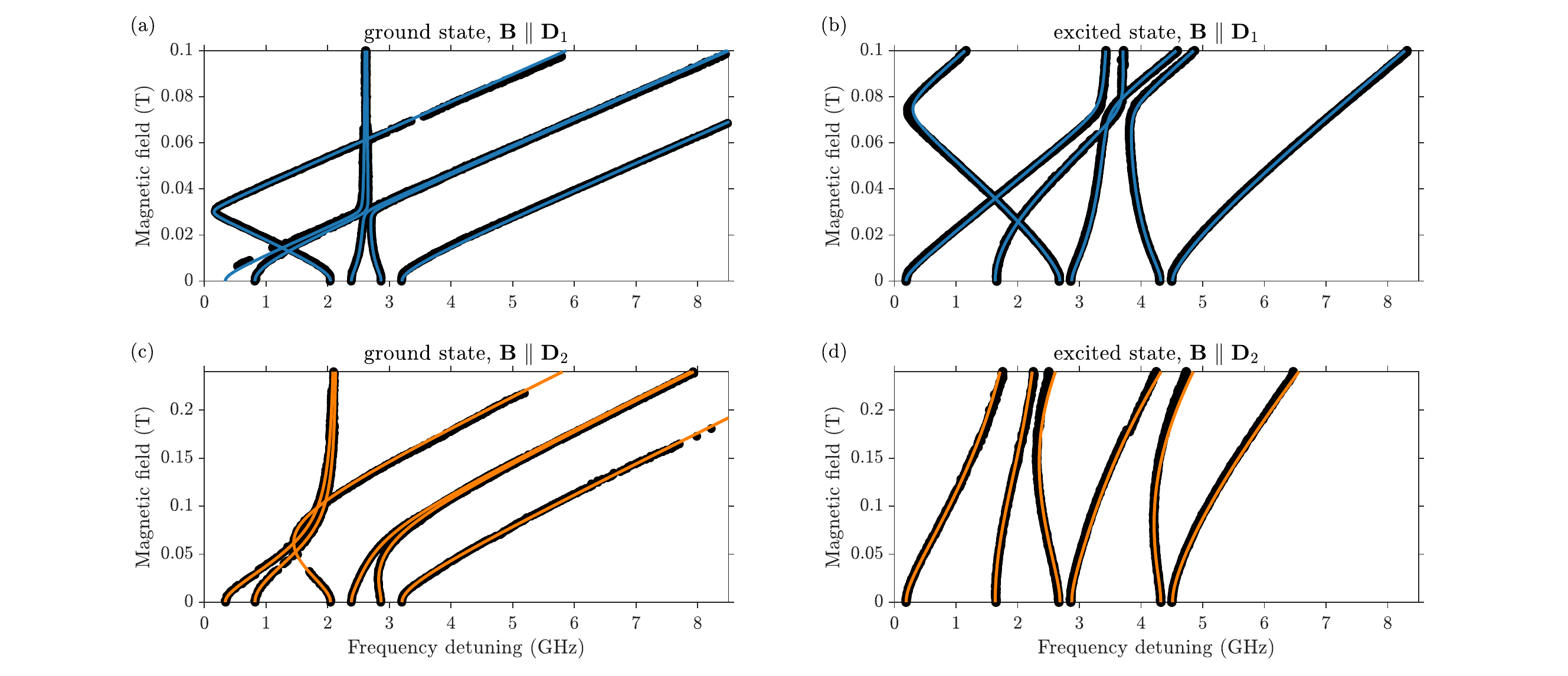}
\caption[]{\label{fig:shb_siteI} (color online) Energies for different spin transitions and fitting results of the hyperfine interaction tensors. 
Transition frequencies for ground (left) and excited (right) states for site~I as a function of magnetic field amplitudes applied in two different directions ($\mathbf{D}_1$ on the top and $\mathbf{D}_2$ for bottom). Experimental points were extracted from SHB measurements (the example is depicted in \figref{fig:map}) and used to fit hyperfine $\mathbf{A}$ tensor parameters (its orientation) given in \tabref{tab:params}. Calculated transition frequencies are plotted with solid lines.
}
\end{figure*}

For optical site~I we detected four ODMR lines at $\left[ 2046,\text{ }2385,\text{ }2869,\text{ }3208 \right]$~MHz (\figref{fig:map}a), while for site~II five ODMR lines were found at $\left[528,\text{ }655,\text{ }2370,\text{ }2496, \text{ }3025 \right]$~MHz (see Appendix, \figref{fig:map2}). \red{The ODMR linewidths varied between 0.5 and 1.2~MHz. These were limited by the inhomogeneous linewidths and possibly by additional power broadening.}

 Using the detected ODMR resonances and Eq.~\eqref{eq:ZFlevels}, one can easily calculate the eigenvalues of the ground-state hyperfine $\mathbf{A}$ tensor. These are listed in \tabref{tab:params} and the corresponding energy level diagrams are reported in~\figref{fig:map}a and~\figref{fig:abs}.
These eigenvalues of the ground-state hyperfine tensor differ considerably from the ones obtained from high field EPR measurements~\cite{Welinski2016}. This clearly demonstrates the problem of standard X-band EPR for characterizing a highly anisotropic magnetic interaction in the case of low site symmetry. The strength of using zero-field ODMR is that we can determine the eigenvalues of the hyperfine tensors, independently of their orientation with respect to the crystal axes and without using any non-linear fitting algorithm. In EPR measurements the eigenvalues and orientation angles (six parameters) are fitted simultaneously to the experimental data, which can introduce errors.

The ODMR technique can also be used to measure hyperfine splittings of the excited state. However, due to the limited excited state lifetime (order of $\approx$1~ms for both sites~\cite{Welinski2016}), excited-state ODMR resonances were not observed in these experiments. The SHB technique was thus used for this purpose. To clearly identify the SHB features stemming from the zero-field hyperfine splittings in the excited state, we found it necessary to also study the SHB spectra as a function of applied magnetic field, as explained in the following.

The SHB pattern appearing inside the absorption profile was recorded at different external magnetic fields in various directions. For each measurement, the normalization signal, that is the absorption profile without hole burning, was subtracted from the acquired trace so to leave only the hole/antihole structure visible \red{(\figref{fig:map}b) (also see Appendix, \figref{fig:map2})}. An example of such a measurement, for site~I and magnetic field parallel to $\mathbf{D}_1$ axis, is shown in~\figref{fig:map}b. It clearly shows a series of lines which shift linearly at high magnetic fields and go into a nonlinear regime at lower magnetic fields.
The holes and antiholes appear as darker and brighter lines, respectively. By following these lines, one can precisely characterize the nonlinear behavior for various magnetic field regimes. \red{The widths of the observed spectral holes were of a few MHz, which were mainly limited by the laser linewidth.}

In conventional SHB methods the side holes usually correspond to the structure of the excited state (see \figref{fig:setup_SHB}).  Therefore, one expects to observe 6 holes on each side of the main hole, corresponding to the 6 possible spin transitions of the excited state. However, in our case we detect a much higher number of side holes, with some of them corresponding to the ground state energy levels. This happens due to the relatively fast relaxation rate on the spin transitions, which is comparable to the burning time used in our experiment (hundreds of milliseconds). In this situation, the optical pumping process effectively empties not only the addressed level but also the ones which are connected to it by fast relaxation processes. This leads to the emergence of pseudo holes at the positions expected for the antiholes, as also explained in~\figref{fig:setup_SHB}c, and it complicates the interpretation of the spectra, in particular the identification of the zero-field hyperfine resonances in the excited state.

In a first step we use the zero-field ground state structure determined from the ODMR data to identify, without ambiguity, several SHB resonances that only depend on the ground state splitting. We use this information to extract the field dependence of the corresponding transitions in a large range of magnetic field amplitudes from the recorded SHB spectra (\figref{fig:map}). The extracted data for site~I is shown in \figref{fig:shb_siteI}.

In a second step we   determine the excited state structure, which requires distinguishing between pseudo-holes and holes. To this end we use the fact that we generally observe two distinct regions of non-linear Zeeman effects in the SHB data. For the particular case of site~I and $\mathbf{B}$ parallel to the $\mathbf{D}_1$ axis, these appear at around 30 and 80~mT, see~\figref{fig:map}b. The one at 30 mT is due to a non-linear Zeeman effect in the ground state, as seen in the already identified  transition energies of the ground state shown in \figref{fig:shb_siteI}a.  The one at 80~mT is then due to the excited state Zeeman effect, which is consistent with the fact that the effective $g$-factor is twice as low in the excited state~\cite{Welinski2016}. Hence, SHB holes that display non-linear behaviour in both the 30 and 80 mT region are pseudo-holes that can be disregarded from the excited state analysis.

In a third step we identify strong SHB side holes, particularly at higher fields, which only display the non-linearity at around 80 mT, as shown in \figref{fig:shb_siteI}b. To further confirm that these are indeed true holes depending on the excited state hyperfine splits, we use the complete set of resonances from both ground and excited states to calculate the position of all expected anti-holes. Many of these are seen in the SHB spectra (see Appendix  \figref{fig:map_SM}), and most importantly none of these predicted lines are in contradiction with the observed SHB spectra. The same approach was also used to analyse the SHB spectra recorded with $\mathbf{B}$ parallel to the $\mathbf{D}_2$ axis for site~I, see \figref{fig:shb_siteI}c-d. We also fully measured and analyzed the SHB spectra along the $\mathbf{D}_1$ and $\mathbf{D}_2$ axes for site~II (see Appendix \figref{fig:shb_siteII}).

The analysis presented above allows us to clearly identify excited-state hyperfine resonances at zero magnetic field, from which the eigenvalues of the excited state hyperfine tensor are calculated. These are listed in \tabref{tab:params} and the corresponding energy level diagrams are reported in~\figref{fig:abs}, both for site~I and II. 

It remains to determine the orientation of the hyperfine $\mathbf{A}$ tensors with respect to the crystal axes. To this end we use the fact that accurate Zeeman $\mathbf{g}$ tensors were already determined ~\cite{Welinski2016} for Yb$^{3+}$ ions having no hyperfine states (I = 0). The hyperfine tensors are thus fitted with respect to the $\mathbf{g}$ tensors, both for the ground and excited states. In the fit we use the hyperfine splittings as a function of magnetic field, measured using the SHB technique. In addition, for the ground state we can use the X-band EPR data from Ref. \cite{Welinski2016} (see Appendix, \figref{fig:epr_siteI}), which was recorded at higher fields in between 100 mT and 1 T. The fitting procedure and the definition of all rotation matrices are given in the Appendix. The fitted rotation angles for all $\mathbf{A}$ tensors are listed in~\tabref{tab:params}. 

\setlength{\tabcolsep}{1.0em} 
{\renewcommand{\arraystretch}{2.0}
\begin{table}[]
\centering
\caption{Hyperfine properties of \ybiso{} on optical \transition{} transition. Principal values of the  $\mathbf{A}$ tensors (in GHz) and Euler angles (in degrees) defining the orientations of tensor's principal axes ($zxz$ convention) in the crystal frame ($\mathbf{D}_1$, $\mathbf{D}_2$, $\mathbf{b}$). The information about Zeeman interaction $\mathbf{g}$ tensors is taken from Ref.~\cite{Welinski2016}.}
\label{tab:params}    
\begin{tabular}{c||cc|cc}    
           & \multicolumn{2}{c|}{Site I} & \multicolumn{2}{c}{Site II} \\ \cline{1-5}
           & ground       & excited      & ground       & excited      \\  \hline \hline
$A_1$ (GHz)      & 0.481       & 1.44         & -0.1259       & 2.34        \\
$A_2$ (GHz)       & 1.159       & 1.82         & 1.1835       & 2.90        \\
$A_3$ (GHz)       & 5.251       & 7.20         & 4.8668       & 6.49        \\
$\alpha_A$ ($^{\circ}$) & 72.25       & 73.88       & 45.86       & 51.07       \\
$\beta_A$ ($^{\circ}$)  & 92.11        & 84.76       & 11.13        & 14.11       \\
$\gamma_A$ ($^{\circ}$) & 63.92       & \red{90.13}       & 2.97       & -0.67     \\   \hline
$\abs{g_1}$      & 0.31       & 0.8         & 0.13       & 1.0        \\
$\abs{g_2}$     & 1.60      & 1.0        & 1.50       & 1.4        \\
$\abs{g_3}$      & 6.53       & 3.4         & 6.06       & 3.3        \\
$\alpha_g$ ($^{\circ}$) & \red{72.8}       & 77       & 59.10       & 54       \\
$\beta_g$ ($^{\circ}$)  & 88.7       & 84       & 11.8        & 23       \\
$\gamma_g$ ($^{\circ}$) & \red{66.2}       &   \red{-7}      & \red{-12.6}       & -10       
\end{tabular}
\end{table}

To determine the order of the  energy splittings we fit the absorption profiles measured at zero magnetic field (\figref{fig:abs}). Only the central frequency offset is used as a free parameter while all the differences between absorption lines are fixed using a certain order of the energy splittings. From the four different possibilities we find that only one gives a spectrum in a good agreement with the experiment. The order found for both sites is depicted in \figref{fig:abs}. This result is further confirmed using separate ODMR measurements by selective optical excitation at the outermost absorption lines. We note that the order is given by the relative signs between the $\mathbf{A}$ tensor eigenvalues, while the simultaneous sign change of two elements does not alter the order. The signs combination corresponding to the determined order is given in \tabref{tab:params} for sites~I and II.

\section{Discussion }
\label{sec:holes}

\begin{figure}
\includegraphics[width=1\linewidth]{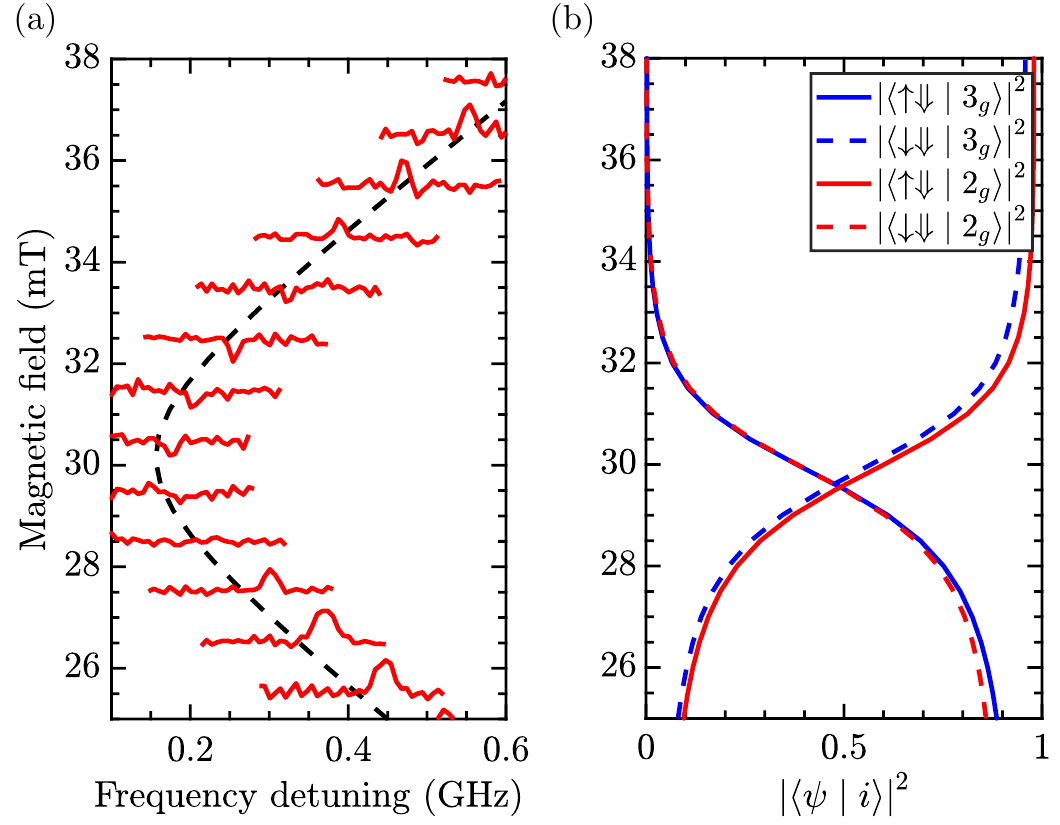}
\caption[]{\label{fig:hole} (color online)
\red{Example of the hole-antihole transformation. (a) Shown is the transformation of the antihole corresponding to the $\ket{2_g}\longleftrightarrow\ket{3_g}$  ground-state transition (at 2046 MHz for zero field). The data is taken from the \figref{fig:map} for magnetic field amplitudes around 30~mT. We believe this transformation to be due to an increased relaxation rate of this transition, which in turn is caused by the rapid change in wavefunctions in this field region (see (b)). The dashed line represents the predicted spectral position of the antihole. The discrepancy between the measured and calculated values can be attributed to the accuracy of the calibration of the laser scan. (b) The change in wavefunction of states $\ket{2_g}$ and $\ket{3_g}$ can be visualized by plotting their overlap with the separable states $\ket{\uparrow \Downarrow}$ and $\ket{\downarrow \Downarrow}$, where $\ket{\uparrow,\downarrow} \equiv \ket{S_z = \frac{1}{2}, -\frac{1}{2}}$ and $\ket{\Downarrow,\Uparrow} \equiv \ket{I_z = \frac{1}{2},- \frac{1}{2}}$. Note that the negligible overlap with the remaining basis states $\ket{\uparrow \Uparrow}$ and $\ket{\downarrow \Uparrow}$ are not shown. All calculations were based on diagonalizing the spin Hamiltonian for the ground state.}}
\end{figure}

A close look at the spectral hole map on \figref{fig:map} (especially in the nonlinear regime around 30~mT) reveals many positions of the antiholes with peculiar behaviour.  As an example in~\figref{fig:hole}a we show  the variation in hole amplitude of the 2046~MHz transition (at zero magnetic field). As seen it features the expected anti-hole behaviour below and above 30~mT, while at around 30~mT it features a pseudo hole. Similar behaviour can be seen also for other lines corresponding to various optical classes (\figref{fig:map}).

We believe this is due to the strong admixing of the corresponding wavefunctions which appears for this magnetic field region. Indeed, from the decomposition of the eigenstates involved in this spin transition (\figref{fig:hole}b) one can see the strong non-linear dependence and flip of the electronic wavefunctions associated with the avoided crossing. This can lead to an enhanced relaxation process on this transition,  which leads to the observed transformation of the antihole into a pseudo hole. \red{We note that a similar hole-antihole transformation was reported in ruby~\cite{Riesen2007}, and the given explanation was also based on the modified cross-relaxation under external magnetic fields.}

The phenomenon of pseduo holes is also observed at zero magnetic field, where the antiholes corresponding to $\ket{1}_g\leftrightarrow \ket{2}_g$ and $\ket{3}_g\leftrightarrow \ket{4}_g$ transitions (823~MHz  and 339~MHz, respectively) appear as pseudo holes. To explain this we assume that the relaxation process at these transitions is faster than for other spin transitions, which makes the optical pumping to be efficient for both of the connected levels (see \figref{fig:setup_SHB}c). The flip-flop relaxation process between different \yb{} ions could be dominant for low magnetic field in this material , as observed in Nd$^{3+}$ doped \yso{}~\cite{ZambriniCruzeiro2017a}, however, further studies of the relaxation rates are required to conclude about the nature of the observed phenomena.

The same behavior is also present for other antiholes on this map, whose energy splittings involve the excited state spin transition, and can be found in many parts of the spectrum. Although this effect complicates the reading of the holeburning spectra, as discussed above, it potentially can also give additional information about the relative relaxation dynamics between particular sets of spin transitions.

Finally we briefly discuss the relative orientations of the $\mathbf{g}$ and $\mathbf{A}$ tensors, and their anisotropy in the ground and excited states. By evaluating the values for hyperfine tensors from \tabref{tab:params} one can see that the orientations of the $\mathbf{g}$ and $\mathbf{A}$ tensors for each state are almost parallel. For site~I the orientation of the maximum $g$-factor is close to $\mathbf{D}_1$, while for site II it is close to $\mathbf{b}$.  The ratio between the tensor elements $A_i/g_i$ is nearly constant for each component. This indicates that each state on the optical \transition{} transition is close to a pure $J$ multiplet~\cite{Welinski2016}. 


\vspace{0.5cm}

\section{CONCLUSION and OUTLOOK}
\label{sec:conclusion}

In conclusion, we have characterized the hyperfine interaction of the \ybiso{} crystal on the optical \transition{} transition. We refined the hyperfine tensor of the ground state and determined its parameters for the excited state. \red{The largest principal values of these tensors are oriented along similar directions for the ground and excited states, which are close to  $\mathbf{D}_1$ axis for site I and close to $\mathbf{b}$ axis for site II. The hyperfine tensors are also similarily oriented as the corresponding Zeeman tensors \cite{Welinski2016}}. This simplifies the description of the spin Hamiltonians of \ybiso{} crystal.

\red{Our characterization of \ybiso{} is in good agreement with previously obtained results for the Zeeman tensors of both the ground and excited states. However, for the hyperfine tensors of the ground states, the eigenvalues are very different as compared to those obtained through conventional EPR measurements in the high-field regime \cite{Welinski2016}. While both hyperfine tensors accurately predict the high-field regime (within experimental errors), only the hyperfine tensors given here work well at low fields. We believe this to be due to the difficulty to accurately determining all eigenvalues in the high-field regime, where the hyperfine interaction is a perturbation to the Zeeman interaction. This problem was also recently encountered in a $^{167}$Er$^{3+}$:\yso{} crystal \cite{Chen2018}, hence it appears to be a general problem that has not yet been studied in EPR studies.}

\red{We also observed pseudo-holes in the SHB spectra, at positions where anti-holes were expected. We attribute this to cross-relaxation processes that efficiently couples certain hyperfine states.} Further studies of the hole lifetimes at different magnetic fields, particularly for the highly non-linear regime, can give the information about relaxation rates and can help to attribute the observed effects to the particular relaxation process.

\red{The spin Hamiltonian presented here allows the calculation of transition frequencies in the spin and optical domain under arbitrarily orientated external magnetic fields, over a wide range of magnetic fields. We note, however, that the effective spin Hamiltonian will break down for very high magnetic fields (for fields significantly higher than 1 Tesla), where perturbations from the nearest crystal-field level cannot be neglected. In particular, the Hamiltonian can be used to accurately predict so-called ZEFOZ (ZEro First-Order Zeeman) points, where the sensitivity of transitions to magnetic field perturbations are highly suppressed \cite{Fraval2004}. In a parallel work the Hamiltonian was used to predict ZEFOZ and near ZEFOZ points in \ybiso{}, at zero and low external magnetic fields, respectively \cite{Ortu2018}. Furthermore, it was experimentally demonstrated that a strong enhancement of both optical and spin coherence times appears in these points, with optical coherence times reaching up to 200~\textmu s and spin coherence times reaching up to 4~ms \cite{Ortu2018}. Based on these results we believe the \ybiso{} crystal is very promising for optical quantum memories \cite{Tittel2010b,Bussieres2013}, microwave-to-optical transducers~\cite{Fernandez-Gonzalvo2015} and coupling to superconducting qubits in the microwave range \cite{Probst2013}.}

\section*{ACKNOWLEDGEMENTS}
The authors thank F.~Fr\"owis and E.~Z.~Cruzeiro for useful discussions, as well as Claudio Barreiro for technical support. We acknowledge funding from EUs H2020 programme under the Marie Sk\l{}odowska-Curie project QCALL (GA 675662), ANR under grant agreement no. 145-CE26-0037-01 (DISCRYS), Nano'K project RECTUS and  the IMTO Cancer AVIESAN (Cancer Plan, C16027HS, MALT).  \\

\vspace{0.5cm}

\bibliography{ytterbium.bbl}

\newpage
\clearpage

\begin{appendix}

\renewcommand{\thefigure}{S\arabic{figure}}

\section{Hamiltonian definition}
\label{app:tensors}

The tensors $\mathbf{A}$ and $\mathbf{g}$ for the ground and excited states can be diagonalized in their respective principle axis systems. To express them in  the crystal frame  we define a rotation with the usual Euler angle convention:
\begin{equation}
\mathbf{A} =R(\alpha_A,\beta_A,\gamma_A)\cdot\begin{bmatrix}
   A_1 & 0 & 0 \\
    0 & A_2 & 0 \\
    0 & 0 & A_3
\end{bmatrix}\cdot R(\alpha_A,\beta_A,\gamma_A)^{T}  
\end{equation}
\begin{equation}
\mathbf{g}=R(\alpha_g,\beta_g,\gamma_g)\cdot \begin{bmatrix}
    g_1 & 0 & 0 \\
    0 & g_2 & 0 \\
    0 & 0 & g_3
\end{bmatrix}\cdot R(\alpha_g,\beta_g,\gamma_g)^{T},
\end{equation}
where $R(\alpha,\beta,\gamma)$ is the rotation matrix with Euler angles $(\alpha,\beta,\gamma)$ for $zxz$ convention
$$R(\alpha,\beta,\gamma) = R_z(\alpha)\cdot R_x(\beta)\cdot R_z(\gamma),$$ where
\begin{align}
R_z(\alpha) = \begin{bmatrix}
    \cos (\alpha) & -\sin (\alpha) & 0 \\
    \sin(\alpha) & \cos (\alpha) & 0 \\
    0 & 0 & 1
\end{bmatrix},
\end{align}
\begin{equation}
R_x(\beta) = \begin{bmatrix}
		1 & 0 & 0 \\
     0 & \cos (\beta) & -\sin (\beta)  \\
      0 & \sin (\beta) & \cos (\beta) 
\end{bmatrix}.
\end{equation}

The interaction tensors for the second magnetic subsite  are defined using an additional $\pi$-rotation around the symmetry $b$  axis of the crystal and given by
\begin{align}
R_z(\pi)\cdot \mathbf{g}\cdot R_z(\pi)^T, \\R_z(\pi)\cdot \mathbf{A} \cdot R_z(\pi)^T.
\end{align}

Additional rotation is applied to go from the crystal $(D_1,D_2,b)$ frame to the laboratory $(X,Y,Z)$ frame and was found to contain the angles lower than 5$^{\circ}$. \red{The orientation of the $\mathbf{D_1}$ and $\mathbf{D_2}$ axes was checked using the polarization dependent absorption measurement. The expected error in their determination could reach 10$^{\circ}$.} The  The interaction tensors in $({D_1,D_2,b})$ crystal frame for the ground and excited state are found to be (in GHz)

\[
\mathbf{A}_{\text{I}}^{(g)} = \begin{pmatrix}
4.847  & -1.232  & -0.244\\ 
   -1.232  &  1.425 &  -0.203\\ 
   -0.244 &  -0.203 &   0.618
\end{pmatrix}_{D_1D_2b} 
\]
\[
\mathbf{A}_{\text{I}}^{(e)} = \begin{pmatrix}
 6.715 &  -1.413  &  0.499 \\
     -1.413 &   2.233  & -0.143\\
    0.499 &  -0.143  &  1.513
\end{pmatrix}_{D_1D_2b} 
\]
   
\[
\mathbf{A}_{\text{II}}^{(g)} = \begin{pmatrix}
   0.686 &  -0.718  &  0.492\\ 
    -0.718 &   0.509  & -0.496\\ 
   0.492   & -0.496 &   4.729
\end{pmatrix}_{D_1D_2b} 
\]
\[
\mathbf{A}_{\text{II}}^{(e)} = \begin{pmatrix}
     2.802  &  -0.379 &   0.661\\
     -0.379  &  2.652  & -0.532\\
    0.661 &  -0.532 &   6.277
\end{pmatrix}_{D_1D_2b} 
\]


The corresponding $\mathbf{g}$-tensors used for the fitting are 

\[
\mathbf{g}_{\text{I}}^{(g)} = \begin{pmatrix}
 6.072  & -1.460  & -0.271\\ 
    -1.460 &   1.845 &  -0.415\\ 
     -0.271 &  -0.415  &  0.523
\end{pmatrix}_{D_1D_2b} 
\]
\[
\mathbf{g}_{\text{I}}^{(e)} = \begin{pmatrix}
  3.242  & -0.566 &   0.249\\
    -0.566  &  0.934  & -0.033\\
      0.249  & -0.033 &   1.023
\end{pmatrix}_{D_1D_2b} 
\]
   
\[
\mathbf{g}_{\text{II}}^{(g)} = \begin{pmatrix}
   0.999 &  -0.766 &  0.825\\ 
     -0.766  &  0.825 &  -0.424\\ 
           0.825 & -0.424 & 5.867
\end{pmatrix}_{D_1D_2b} 
\]

\[
\mathbf{g}_{\text{II}}^{(e)} = \begin{pmatrix}
     1.389  & -0.337  &  0.572\\
     -0.337  &  1.308 &  -0.383\\
      0.572 &  -0.383 &   3.008
\end{pmatrix}_{D_1D_2b} 
\]

\begin{figure*}
	\includegraphics[width=0.83\linewidth,trim=0.4cm 0.0cm 0cm 0.5cm,clip] {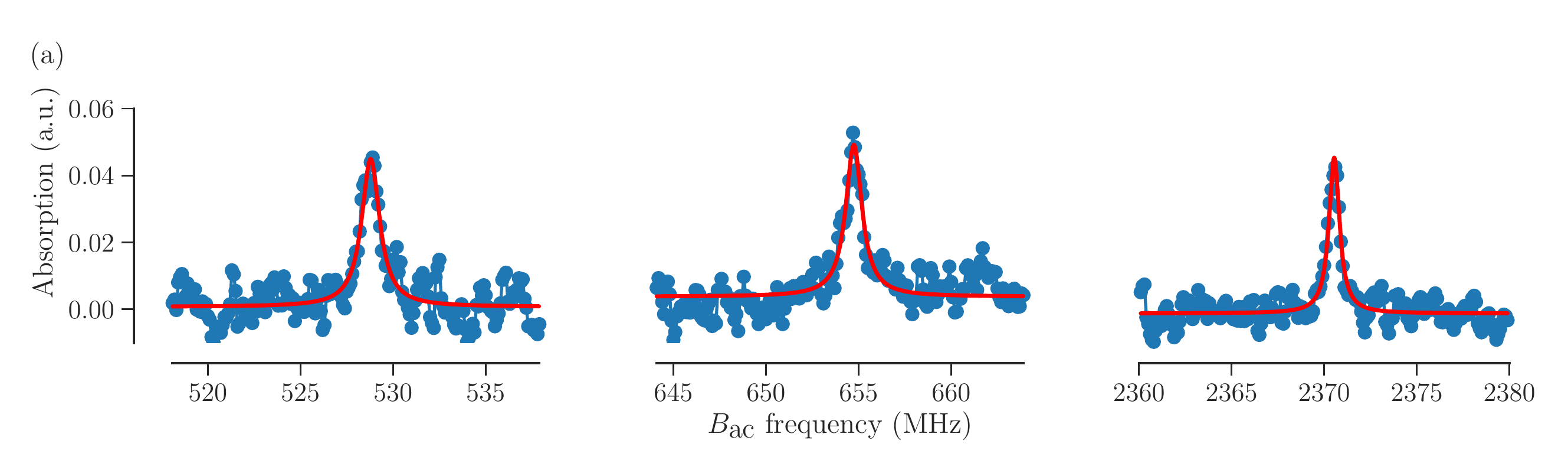}\includegraphics[width=0.17\linewidth,trim=0.0cm -0.9cm 0cm 0cm,clip] {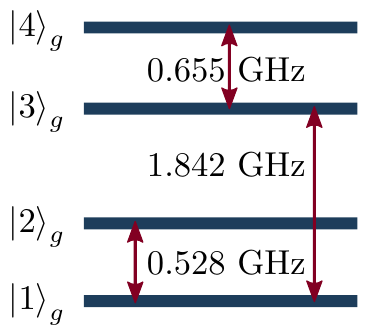}\\
	\includegraphics[width=1.0\linewidth,trim=2.3cm .0cm 4cm 0.cm,clip] {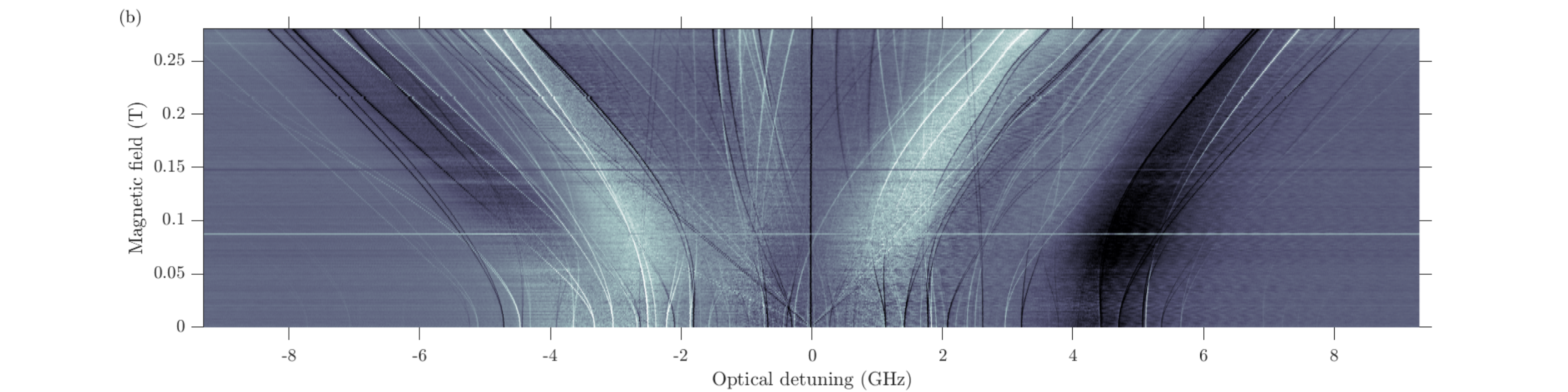} 
	\includegraphics[width=1.0\linewidth,trim=2.3cm .0cm 4cm 0.cm,clip] {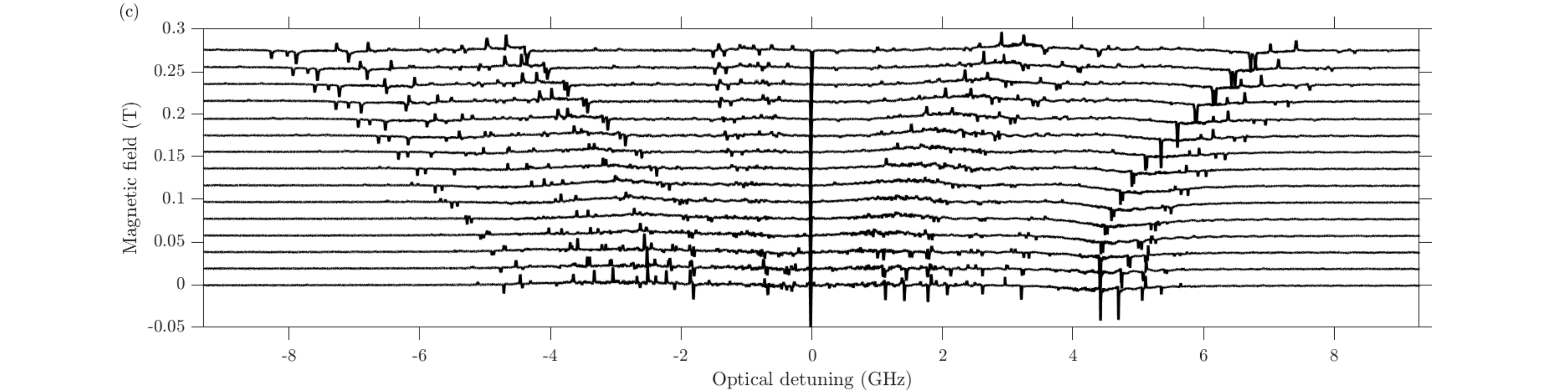}
\caption[]{\label{fig:map2} (color online) Experimental results. 
(a) Optically detected magnetic resonance (ODMR) signals measured for site II and the reconstructed energy level structure of the ground state. The oscillating magnetic field $B_{\text{ac}}$ was applied in $\mathbf{b}$ direction.  Three different transitions are shown and their corresponding levels are depicted on the right.
(b) Recorded optical spectral holeburning (SHB) spectra for site~II of \ybiso{} crystal measured for different magnetic field amplitudes applied in the direction close to $\mathbf{D}_2$-axis. Zero frequency detuning corresponds to the central frequency at which spectral holeburning is performed. Black regions correspond to lower absorption (holes), while white lines correspond to lower transmission (antiholes) regions. White and black dashed lines indicate energy level splittings of ground and excited states, respectively.  
(c) The examples of SHB spectra taken from (b) with lower for magnetic field resolution.
}
\end{figure*}

\begin{figure*}
\includegraphics[width=1.0\linewidth,trim=1.6cm 0.1cm 2.6cm 0.05cm,clip] {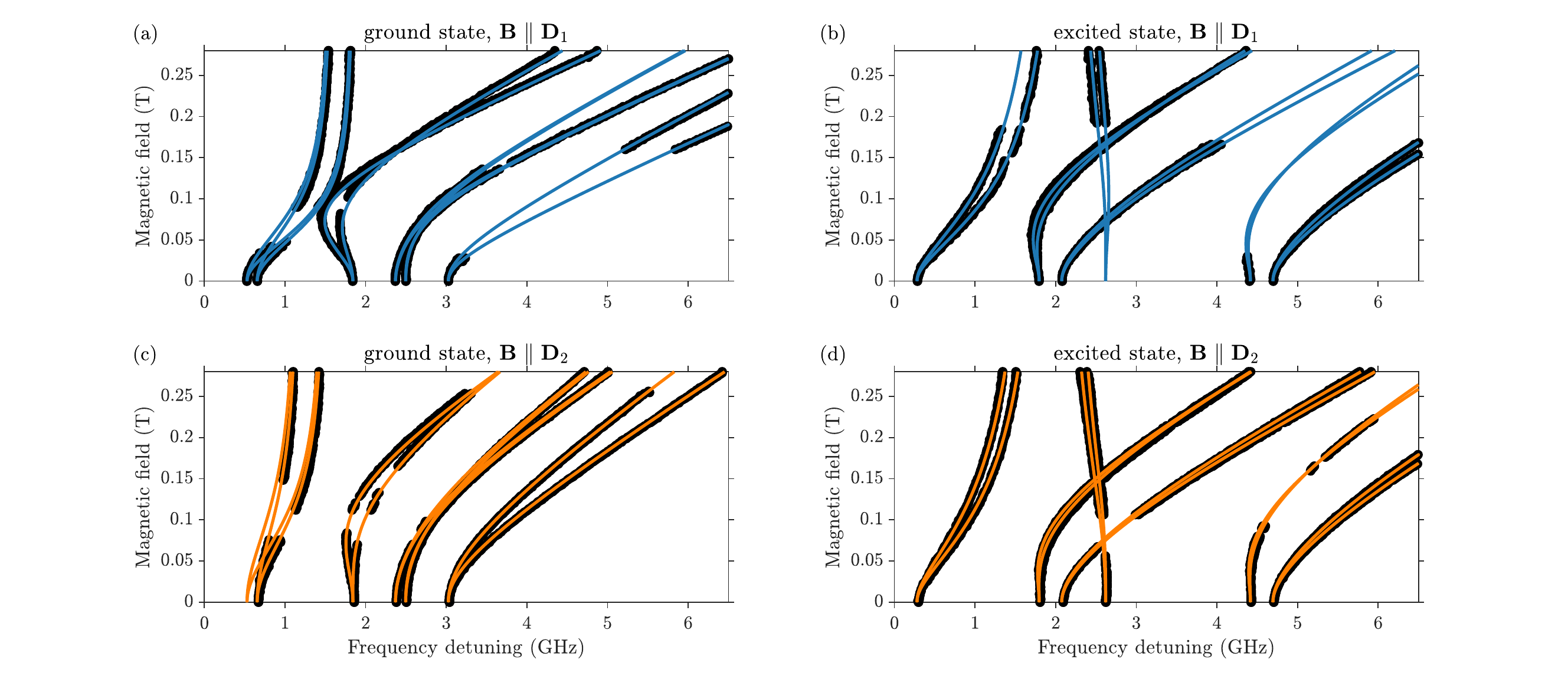}
\caption[]{\label{fig:shb_siteII} (color online) 
Transition frequencies for ground (left) and excited (right) states for site~II as a function of magnetic field amplitudes applied in two different directions ($D_1$ on the top and $D_2$ for bottom). Experimental points were extracted from SHB measurements (for example \figref{fig:map}) and used to fit hyperfine $\mathbf{A}$ tensor parameters given in \tabref{tab:params}. Calculated transition frequencies are plotted with solid lines.
}
\end{figure*}

\begin{figure*}
\includegraphics[width=0.49\linewidth,trim= 0cm 1.2cm 0cm 1.2cm] {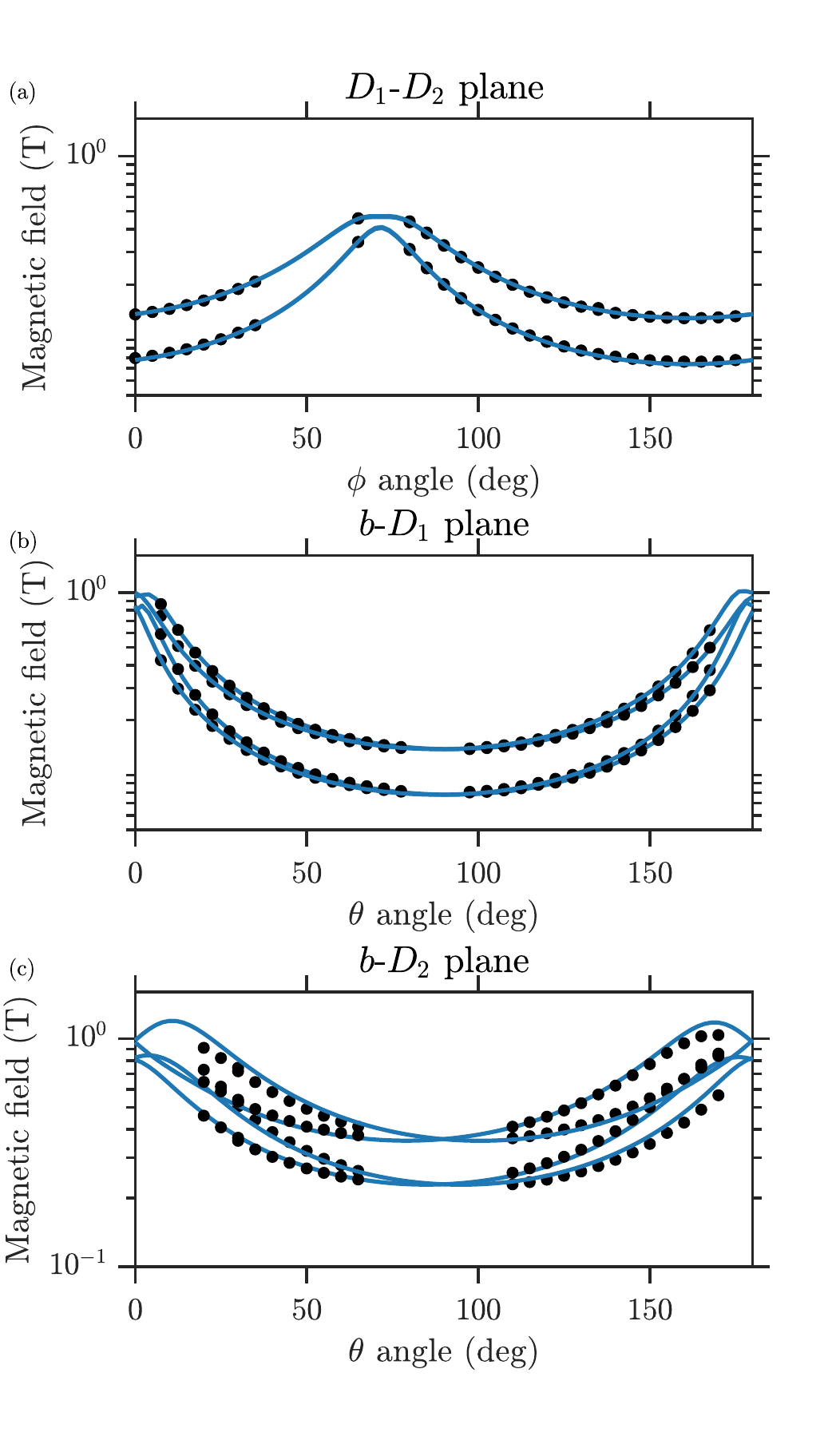}
\includegraphics[width=0.49\linewidth,trim= 0cm 1.2cm 0cm 1.2cm] {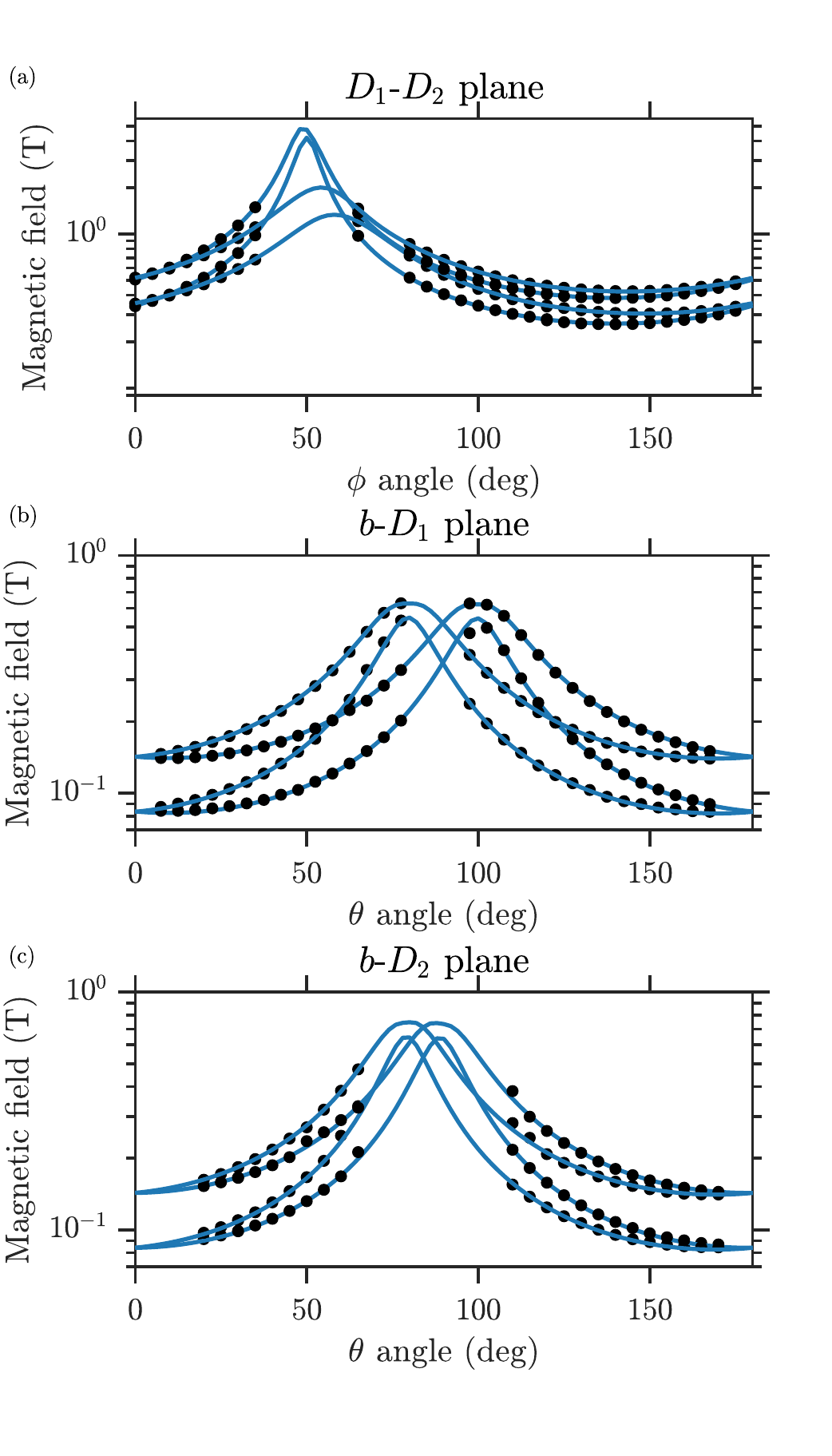}
\caption[]{\label{fig:epr_siteI} (color online)
Angular variations of the EPR transitions of \ybiso{} ground state for site~I (left) and site~II (right) in the three perpendicular crystallographic planes ($D_1-D_2$, $b-D_1$ and $b-D_2$) . Experimental data (points) are well described by the model based on previously measured $\mathbf{g}$-tensors \cite{Welinski2016} and fitted hyperfine $\mathbf{A}$ tensors in this work (solid lines).
}
\end{figure*}

 \clearpage
 \newpage
\begin{turnpage}
\begin{figure*}
   \includegraphics[width=1.\linewidth,trim=2.6cm 0.5cm 2.7cm 2cm,clip] {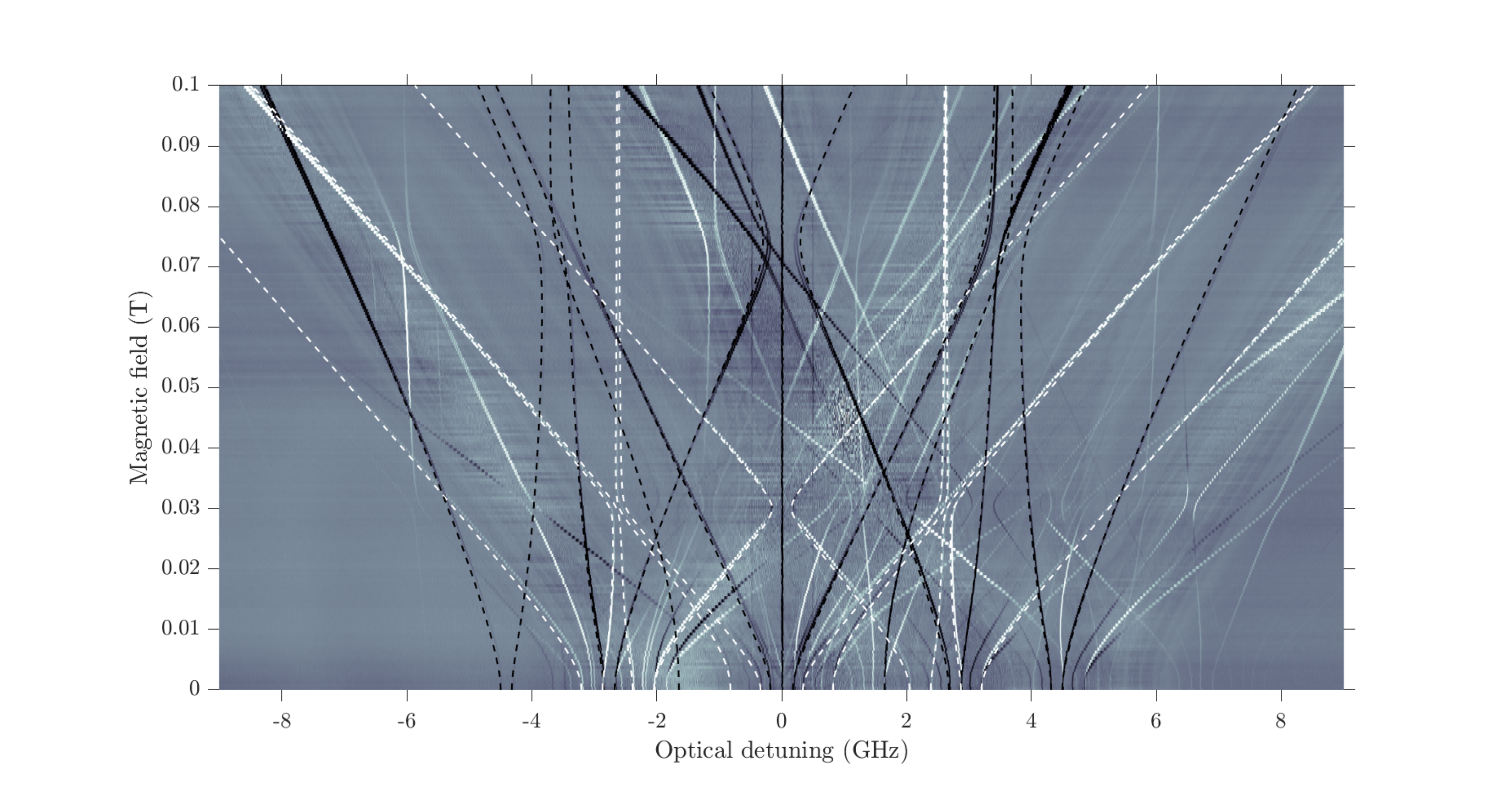} 
\caption[]{    \label{fig:map_SM} (color online)
Recorded optical spectral holeburning  spectra for site~II of \ybiso{} crystal measured for different magnetic field amplitudes applied in the direction close to $\mathbf{D}_2$-axis. Zero frequency detuning corresponds to the central frequency at which spectral holeburning is performed. Black regions correspond to lower absorption (holes), while white lines correspond to lower transmission (antiholes) regions. White and black dashed lines indicate energy level splittings of ground and excited states, respectively.  
\red{The deviation from the data is attributed to the imperfect calibration of the laser scan.}
}
\end{figure*}


\end{turnpage}

\end{appendix}


\end{document}